# Digital Frequency Domain Multiplexer for mm-Wavelength Telescopes

Matt Dobbs, Eric Bissonnette, and Helmuth Spieler

*Abstract*— An FPGA based digital signal processing (DSP) system for biasing and reading out multiplexed bolometric detectors for mm-wavelength telescopes is presented. This readout system is being deployed for balloon-borne and ground based cosmology experiments with the primary goal of measuring the signature of inflation with the Cosmic Microwave Background Radiation. The system consists of analog superconducting electronics running at 250mK and 4K, coupled to digital room temperature backend electronics described here. The digital electronics perform the real time functionality with DSP algorithms implemented in firmware. A soft embedded processor provides all of the slow housekeeping control and communications. Each board in the system synthesizes multi-frequency combs of 8 to 32 carriers in the MHz band to bias the detectors. After the carriers have been modulated with the sky-signal by the detectors, the same boards digitize the comb directly. The carriers are mixed down to base-band and low pass filtered. The signal bandwidth of 0.050 Hz – 100 Hz places extreme requirements on stability and requires powerful filtering techniques to recover the sky-signal from the MHz carriers.

## I. INTRODUCTION

A NEW generation of mm-wavelength instruments is being commissioned including APEX-SZ [1], The South Pole Telescope (SPT) [2], and the Atacama Pathfinder Telescope (ACT) [3]. These instruments employ hundreds or thousands of bolometric detectors to accurately measure the Cosmic Microwave Background (CMB) radiation. The improved sensitivity of these instruments is made possible by new detector and readout technology that is easily scalable to large focal plane arrays. A key element of this technology is SQUID-based multiplexed readout systems. Multiplexing the detector readout greatly reduces the heat load due to wiring on the detector cold-stage, reduces the complexity of cold components in the system, and reduces cost. Together with our collaborators at UC Berkeley, we developed the frequency domain multiplexed (fMUX) system [4], [5] that is being used for APEX-SZ, SPT, and elsewhere. The group at NIST and UBC have developed a time domain multiplexed system [6], [7] that is being used for ACT and elsewhere. Both systems achieve their basic performance requirement—the detectors are read out continuously without any appreciable increase in noise due to the readout system. These systems are designed for ground based instruments. The next technological step is to deploy the systems for space telescope applications onboard stratospheric balloon or satellite missions. The fMUX system currently draws too much power for these applications. In the course of the analog system's development, fast ADCs and FPGAs with substantially increased logic and reduced power consumption became available. This allowed for the development of digital backend electronics for the fMUX system that dissipate an order of magnitude less power. The system has advantages in its configurability and its parallel nature. The digital fMUX system will be flown on the EBEX balloon-borne CMB polarimeter [8] next year and is planned for use on the forthcoming POLARBEAR [9] instrument and the proposed polarization upgrade to the SPT. These experiments will measure the faint polarization signature of the CMB revealing or providing exclusion limits on the faint signature of gravity waves produced during inflation.

The digital frequency domain multiplexer (DfMUX) is designed to read out many Transition Edge Sensor bolometers on a single set of wires without appreciably contributing to the system noise. The detectors are low impedance (≈ ½ Ω) devices cooled to 0.25K with a noise level of several $10^{-17}$ W/√Hz. The sky signals are modulated in the bandwidth 0.05-100 Hz by the motion of the telescope or optics. The low noise and low impedance requirements are ideally matched to Superconducting Quantum Interference Devices (SQUIDs) for cold pre-amplification. The low frequency noise specification places strict requirements on all aspects of the system and distinguishes the firmware and digital algorithms employed in this system from other modulation/demodulation applications such as software defined radio.

The electronics system also needs to tune the detectors to the optimum bias point by adjusting their voltage bias and tune the SQUID pre-amplifiers using bias currents to obtain the best noise performance and dynamic range. Providing these control functions is non-trivial and few other applications have required the level of automation that is

Manuscript received May 11, 2007. This work was supported in part by the National Science and Engineering Research Council of Canada. We thank Xilinx University Programs Canada for in-kind hardware contributions.

Matt Dobbs is with McGill University, Rutherford Physics Building, 3600 University Street, Montréal, QC H3A 2T8 (phone: 514-398-6500; e-mail: Matt.Dobbs@McGill.ca).

Eric Bissonnette is with McGill University, Rutherford Physics Building, 3600 University Street, Montréal, QC H3A 2T8 (e-mail: Bissonnette@Astro.UMontreal.ca).

Helmuth Spieler is with Physics Division, Lawrence Berkeley National Laboratory, 1 Cyclotron Road, Berkeley, CA 94720 (e-mail: HGSpieler@LBL.gov).

necessary for this remote balloon telescope with thousands of channels.

## II. SYSTEM OVERVIEW

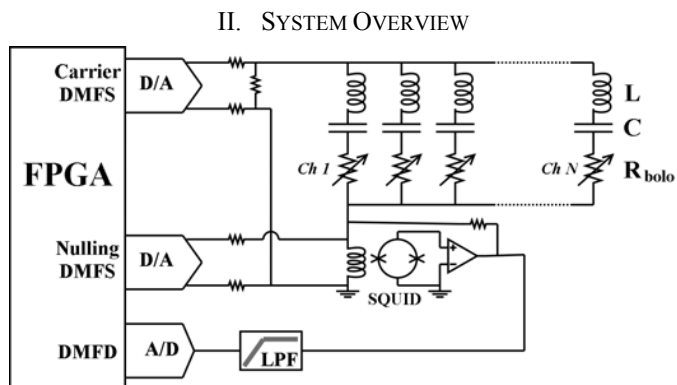

Figure 1 The digital frequency domain multiplexer system is shown schematically.

The digital frequency domain multiplexer (DfMUX) is shown schematically in **Figure 1**. The bolometer sensors $R_{bolo}$ are biased with sinusoidal voltages in the frequency range from 300 kHz to 1 MHz. Each bolometer is biased at a different frequency. Intensity variations from the sky-signal change the bolometer resistance and amplitude modulate the bolometer current such that the sky-signal from each bolometer is transferred to a sideband adjacent to its carrier. Thus, the signals from different bolometers within a module are uniquely positioned in frequency space, so they can be summed and connected through a single wire to a SQUID preamplifier operating at 4K. Each bolometer is connected through a series resonant LC circuit that defines the bias frequency. This allows the bias frequencies for all bolometers in a module to be applied through a single wire, as the tuned circuit selects the appropriate frequency for each bolometer. Only two wires are needed to connect the bolometers of a readout module on the 0.25K stage to the 4K stage on which the SQUIDs are mounted. The comb of bias carriers is synthesized with a *Digital Multi-Frequency Synthesizer* (DMFS) described in Section III. The tuned circuits also limit the bandwidth of the bolometer Johnson noise, which would otherwise contribute to the noise in all other channels of the module.

The SQUID is the only device we know of that has the necessary noise temperature and low input impedance to satisfy this system's requirements. However, the output of a SQUID amplifier is periodic vs. input signal, so input signals must be limited to a monotonic portion of the roughly sinusoidal output. Since the carrier amplitudes are orders of magnitude larger than the sky signals, we cancel the carriers at the SQUID input with a second comb (synthesized with a second DMFS), referred to as the nulling signal. This nulling comb is an inverted version of the original carrier comb, and serves to remove the large carrier signals. It does not affect the carrier sidebands, which contain the information of the sky-signals. The use of the nulling signal dramatically reduces the dynamic range requirements of the system. Nulling factors of $10^3$ are routinely achieved.

The SQUID amplifiers are 100 element series array SQUIDs [10] manufactured by NIST in Colorado. Each SQUID device is operated in shunt-feedback with a low-noise bipolar transistor op-amp located on a custom room temperature SQUID controller circuit board. These boards also include digital control electronics and DACs to provide the bias currents and tuning functionality to operate the SQUID devices.

The SQUID controller and cold components of the system were developed by a collaboration at LBNL/UC Berkeley (including two of the authors) and are described in [4], [5].

After amplification, the comb of amplitude modulated carriers output by the SQUID controller is transmitted to the Digital Multi-Frequency Demodulator (DMFD) where the comb of signals are directly digitized and processed as described in Section IV.

Each module of the digital backend electronics— encompassing a carrier DMFS, nulling DMFS, and DMFD— is capable of reading out a module of 32 bolometers. Presently the SQUID controller bandwidth limits the system to 12 bolometers per module, as will be deployed for the EBEX instrument. Improvements to the SQUID system are being developed in collaboration with UC Berkeley/LBNL that could dramatically increase the number of bolometers per module.

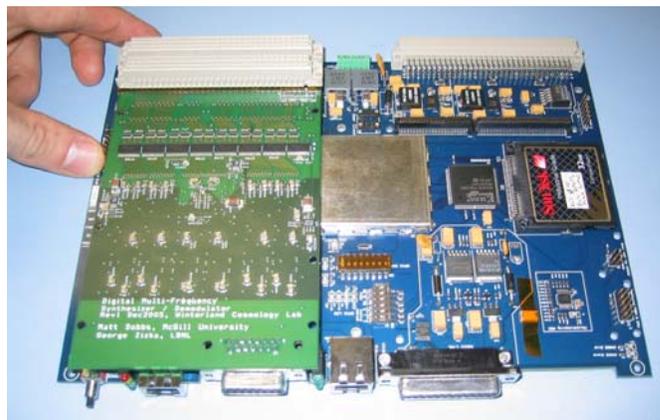

Figure 2 An FPGA motherboard is shown with one of its two analog converter mezzanine boards attached.

Four DfMUX modules are contained on a single 6U VME board, shown in Figure 2. Each motherboard can handle 128 bolometer channels (48 channels with the 12 bolometer multiplexing scheme). The rather complex digital circuits, including a powerful Xilinx Virtex4 LX100 FPGA, reside on this FPGA motherboard, while the low noise analog converters, amplifiers, and filters ride 'piggy-back' on two mezzanine boards. A SQUID controller board, capable of handling 8 SQUID modules, forms an intermediary between the cold SQUIDs and the analog signals coming from the DfMUX backend. Thus the 1536 detector channel EBEX balloon telescope instrument has 32 DfMUX backend boards, 16 SQUID controller boards, and 128 SQUIDs. The SQUID controllers are mounted in a Faraday cage directly on the receiver cryostat. The DfMUX boards are housed in two 6U VME racks.

The SQUID devices are very sensitive to pickup of fast digital signals and so the isolation between the digital components, analog circuitry, and SQUID controllers is crucial. The digital components of the FPGA motherboard are placed inside a Faraday shield formed by a plane layer in the printed circuit board and a metallic surface mount cover visible in the center of Figure 2. The analog components of each Mezzanine board are placed inside similar shields and the ground connections between the Mezzanines and motherboard are carefully chosen.

## III. DIGITAL MULTI-FREQUENCY SYNTHESIZER

The Digital Multi-Frequency Synthesizer (DMFS) produces a comb of up to 32 sine waves using an algorithm implemented in firmware. It converts the signal to analog using a 16-bit D/A operating at 25 MHz.

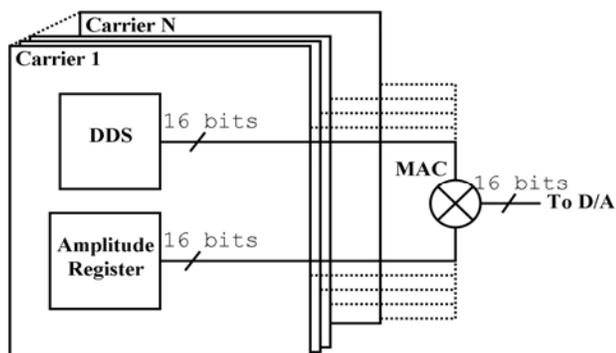

Figure 3 The Digital Multi-Frequency Synthesizer algorithm is shown schematically.

The DMFS firmware is shown schematically in **Figure 33**. Each sinusoidal carrier is synthesized using a Direct Digital Synthesizer algorithm [11]. This algorithm uses a 32-bit phase accumulator register to keep track of the waveform's phase. Each clock cycle this accumulator is incremented and truncated to 16-bits then used to reference the address of the waveform's 16-bit amplitude in a look-up table. A separate 32 bit register provides a programmable phase offset. When operating at 25 MHz, the frequency can be specified to 0.006 Hz and the spurious free dynamic range (SFDR) is 96 dB. This algorithm does not use substantial logic resources or power, but does use significant memory (~1 Mbit). By multiplexing the look-up table for 8 sine waves, the memory usage is minimized. Each of the sine waves is multiplied with a 16 bit amplitude adjustment register, truncated to 18-bits, and accumulated over all carriers in the comb. The result is truncated to 16-bits and sent as a parallel LVCMOS signal to a 16 bit D/A converter on the mezzanine board. Clock bleed-through is removed from the analog output signal with a passive 4-pole low pass filter operating at 8 MHz. A differential amplifier with four selectable gain settings allows for large changes in bias voltage without compromising the digital dynamic range. The comb is sent to the cryostat as a differential signal on a shielded twisted pair cable.

The stability of the DMFS waveforms is extremely important due to the low frequency noise requirements of this system. The sinusoidal carriers are 'bucked' by the sinusoidal nulling waveforms. Though this subtracts away the coherent portion of these waveforms, the low frequency noise in these waveforms will add incoherently, contaminating the sky-signals. The DMFS carrier, DMFS nuller, and DMFD are all clocked with the same crystal oscillator so that clock jitter cancels out to first order. The largest contributor is the low frequency noise from the transistors in the D/A converter output ladder. This low frequency transistor noise is modulated up to the carrier tone frequency by the D/A switching and appears as sidebands on the tones. This parameter is rarely optimized in modern D/A devices and is typically not shown in datasheets. To find a suitable commercial device, D/A converters with large transistor gate area (typically older devices) and adequate speed were selected. Test circuits were built for four devices and low frequency noise measurements performed. The system requirement is the sideband noise to carrier ratio 1 Hz away from the carrier should be better than $10^{-6}/\sqrt{Hz}$. The frequency spectra for these devices are shown in **Figure 44**. Two devices satisfy our criteria and we chose the device with the lowest power consumption, the Linear Technology LTC1668 for use in the DMFS.

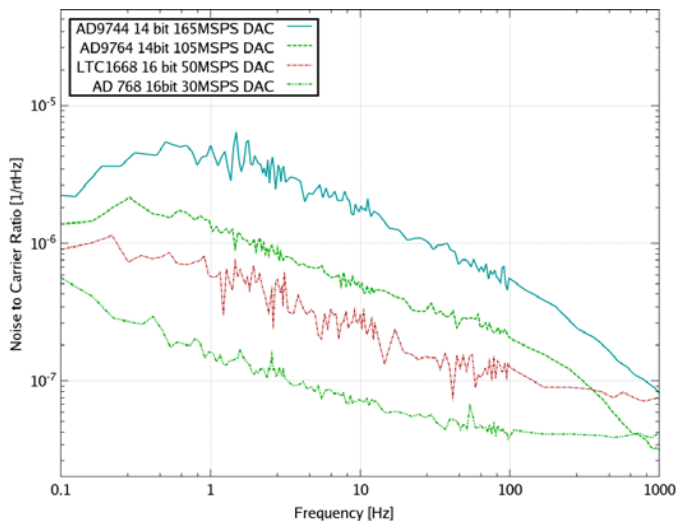

Figure 4 Noise to carrier ratio of four commercially available D/A devices. Two of these devices satisfy our requirement of $<10^{-6}/\sqrt{Hz}$.

## IV. DIGITAL MULTI-FREQUENCY DEMODULATOR

The output of the SQUID controller is a frequency comb modulated by the sky-signal. At this stage the carriers are relatively small due to the nulling described above. This comb is transmitted on a twisted pair cable to the Digital Multi-Frequency Demodulator (DMFD), which digitizes the waveform, mixes the individual sky-signals down to base-band, and applies a low pass filter.

The analog portion of the DMFD consists of a differential amplifier with four selectable gain settings, a 4 pole passive low pass anti-aliasing filter operating at 8 MHz, and a 14-bit A/D converter operating at 25 MHz. These components are housed on the mezzanine board and the digital signals are transmitted as LVCMOS signals to the FPGA. The over-sampling improves the resolution beyond 14 bits. Low frequency noise in the A/D converter's transistors is typically not an issue, because their noise level is normally well below the level of the least significant bit.

Since the sky-signals occupy only a small fraction (about 1 part in 1000) of the waveform's total bandwidth, it is not feasible to store the entire waveform on computer disk as the required disk space would be increased by a factor 1000. The signals need to be processed in real time.

Inside the FPGA, the DMFD input data is sent down a set of parallel algorithm pipelines, each of which consists of a quadrature mixer, Cascade Integrator-Comb (CIC) low pass filter, and a chain of band defining FIR filters. There is one pipeline for each detector channel in the comb. A basic schematic of the configuration is shown in

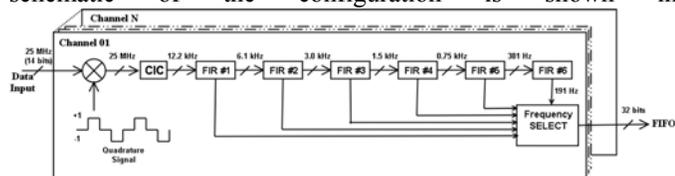

Figure 5 5.

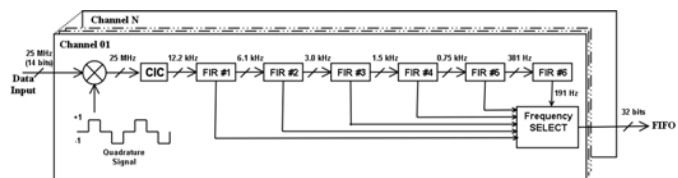

Figure 5 The Digital Multi-Frequency Demodulator algorithm is shown schematically.

The first component in the digital demodulator is a square wave quadrature mixer that down-converts the sky signal to base-band by applying a factor +1, -1 or zero to the waveform. The square wave mixer requires far less memory and logic resources than a sine wave algorithm because it performs down-conversion without multiplication. The mixer uses 32-bit phase accumulator and phase offset registers such that the mixer is locked at the same frequency as the DDS that generates the carrier sine wave, but can account for arbitrary phase shifts (arising in the cryogenic wiring) with the offset register.

The low pass filtering of the waveform is challenging. Low frequency signals need to be maintained while filtering and sub-sampling the waveforms by a factor of roughly $10^5$ to reduce the 25 MHz sampling rate to the ~200 Hz data rate that will be recorded on disk.

The first stage of the digital filter is a CIC or Hogenauer filter [12] providing a factor 2048 of decimation, resulting in an output frequency of 12.2 kHz. CIC filters are implemented using only adders, subtracters, registers, and delays. There are no multiplication operations in the algorithm, rendering it ideal for fast parallel FPGA applications. The CIC is best used for digital down conversion when there is a fast input data stream with the signal occupying only a small fraction of the bandwidth. This is the case for the multiplexer application, since the bolometer information is contained in a narrow bandwidth of about a hundred Hz around the carrier. An N-stage CIC filter is equivalent to a cascade of N boxcar filters. The draw-back to a CIC filter is that it has considerable 'droop' in the pass-band. For this CIC filter, the decimated output would have a 12 kHz sample rate. The droop across the full 12 kHz band is tremendous, but of no concern, since the sky signal lies in a narrow band (typically 0.05-100~Hz) where the droop is small. Since the CIC is effectively integrating a large number of timestream samples, there is rapid growth in the maximum number of bits, and the output is 81 bits.

Following the CIC filter and convergent truncation to an 18-bit signed integer, a series of several FIR filters with a top-hat shape each decimate by a factor of 2, further narrowing the bandwidth. The computational usage of these filters is negligible, since they operate at slow speed, and so we can use a large number of taps per filter. The present configuration uses 128 taps, and produces 120dB of stop-band attenuation. The most costly operation in an FIR filter is the multiply-and-accumulate (MAC) function for which we use dedicated MAC components on the Xilinx FPGA device. The user can easily change the output data rate and hence the bandwidth of the system by bypassing some of the FIR filters with simple software commands. This is a powerful tool for debugging during the systems integration phase of the instrument's commissioning, where typically a larger bandwidth is desired to measure detector properties such as electrical and thermal time constants.

The output of the low pass filter is sent to First-In-First-Out (FIFO) memory. The embedded soft processor retrieves the output from the FIFO, packages it together with data from the three other DFMD modules on the FPGA motherboard, and sends the data as User Datagram Protocol (UDP) packets over Ethernet.

V. THE FPGA MOTHERBOARD

The FPGA motherboard hosts two mezzanine boards, each of which contains the analog converters for two multiplexer modules. A powerful Xilinx Virtex4 LX100 or LX160 FPGA is the heart of the motherboard. The FPGA is interfaced to 64 MB of external DDR memory and a 10/100/1000 Mbit Physical layer device providing a gigabit Ethernet connection.

The FPGA includes the real-time digital signal processing, implemented in firmware, and an embedded soft processor (also implemented in firmware) running the Xilinx µBlaze operating system. Each of the four multiplexer firmware modules includes a carrier DMFS, nuller DMFS, and DMFD which feeds a FIFO. Software, written in C for the µBlaze processor, provides all of the slow control functionality and interface protocol. The software includes a web-server that is accessible via html and provides the user with web-based forms that allow for the configuration of the system. This html interface protocol is normally employed when using the board to test and evaluate a small number of detectors. In this mode, the electronics can be controlled from any computer with web access, and no custom software is needed. When many FPGA motherboards are used in a larger system, they are configured from a control computer using TCP/IP commands. In both cases, the embedded processor assembles detector timestreams from the FIFO and transfers them to an external data acquisition computer using UDP packets transmitted over Ethernet.

An advanced set of tuning algorithms, also written in C, optimize the tuning for the SQUID and bolometer devices.

### A. Configuration and clock

Configuration bitstreams and disk space are provided by a compact flash card. This format is familiar to most users, making it easy for them to install firmware upgrades. A large 8 GB compact flash card has enough storage capacity for data from several days of continuous detector readout, providing a good backup system for data that is normally stored on a central control computer.

The motherboard has an internal crystal oscillator clock that is used when the board runs in stand-alone mode. When running in a larger system with other motherboards, the clock is produced by one master board (selected with an external jumper) and distributed as an LVDS signal across the backplane to the other boards such that all FPGA motherboards in the system share a single clock. For synchronization with other telescope subsystems, a timestamp can be input on the front panel as a differential RS485 signal. This timestamp (normally derived from GPS) can be Manchester encoded or formatted as a pulse width modulated (PWM) IRIG-B signal.

### B. Housekeeping and slow controls

Temperature and voltage monitoring is important for these boards as they will be subject to extreme conditions as the EBEX balloon ascends through the atmosphere. The temperature sensors allow the electronics to be automatically shut down if the boards overheat or if extreme exterior temperatures are encountered. The temperature is measured at five locations on each board. All of the pre-regulator input voltages are monitored with an A/D converter, allowing for tracking of external voltages supplied by batteries and solar panels.

The SQUID controller boards are interfaced through a DB25 connector on the front-panel which provides power and digital control commands via LVDS twisted pairs.

A watchdog circuit on the board will issue a reset command to the embedded soft processor if the processor becomes unresponsive.

### C. Power and thermal considerations

The board receives unregulated power through the backplane. Three PWM buck regulators on the FPGA motherboard provide power (3.3V, 2.5V, 1.8V) for the digital components. A separate set of three linear regulators on each mezzanine board provides power for the sensitive analog components (±5V, 3.3V).

The power consumption of the FPGA motherboard is about 6W and each mezzanine board draws 4W. The SQUID controller board draws about 4W. The power consumption per multiplexer module is thus about 4W, representing an order of magnitude improvement over our previous analog fMUX system [1], [2]. The DfMUX readout system for the 1536 channel EBEX balloon instrument, multiplexing 12 detectors per module, will draw a total of about 500W. Further improvements in power consumption will be realized by increasing the multiplexing factor from 12 up to 32. This will require developments that extend the bandwidth of the SQUID controller and is an issue that is currently being addressed.

Cooling for electronics on-board high altitude balloon flights is challenging. The low atmospheric pressure means that convective cooling is negligible. The DfMUX electronics are designed to be cooled conductively, avoiding the use of a pressure vessel. All of the plane layers in the 14-layer FPGA motherboard and the 6-layer Mezzanine board use extra thick copper, to improve thermal conductivity. The FPGA device and Mezzanine boards are thermally connected to high thermal conductivity alloys (not shown in Figure 2) that are thermally sunk to the instruments aluminum gondola. Power dissipated to the gondola is radiated towards the ground. The thermal concept for the EBEX instrument is presently undergoing testing.

## VI. SYSTEM NOISE PERFORMANCE

An early prototype of the system has been tested end-to-end with transition edge sensor detectors using a test cryostat at UC Berkeley. This test cryostat had five detectors connected in a frequency domain multiplexed configuration. The detectors were biased into their superconducting transition and noise spectra were recorded (**Figure 11**Figure 6). These spectra are consistent with the expectation from the detectors alone. the noise contribution of the SQUID and readout electronics (superimposed in Figure 6) contribute negligibly to the system noise.

The DfMUX electronics for the test-flight of the EBEX instrument, scheduled for early summer 2008, have been constructed and are being integrated now with the detectors system. Performance verification and commissioning results will be available as the integration progresses.

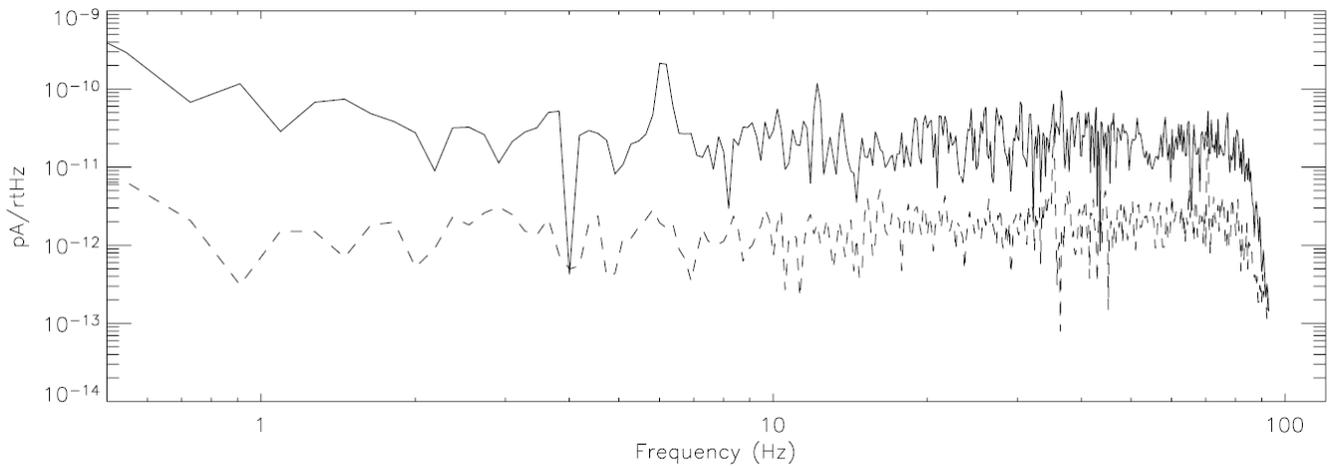

Figure 6 Noise power spectral density for a transition edge sensor bolometer read out with the digital multiplexer system is shown (solid line). The white noise is consistent with the detector noise expectation of 20 pA/√Hz. The peak at 6 Hz is due to a feedback loop that is stabilizing the detector temperature. The noise spectrum for the readout system alone is superimposed (dashed line).

## VII. CONCLUSION

Recent developments in the processing power of FPGAs have allowed for the development of digital backend electronics for a SQUID-based frequency domain multiplexer system that operates with large arrays of sub-Kelvin Transition Edge Sensor bolometers. This new technology has sufficiently low power consumption to allow for the readout of large focal plane arrays on stratospheric balloon platforms. The system will be deployed on the EBEX instrument in 2008.

This system addresses a unique combination of challenges: (1) maintaining low noise figure (several $10^{-17}$ W/√Hz) of the detectors, (2) maintaining sky-signals at low ($\leq$ 0.1 Hz) frequency, (3) necessity of filtering and sub-sampling thousands of detector time-streams by a factor of $10^5$, and (4) susceptibility of the SQUIDs to fast digital signals.

## ACKNOWLEDGMENT

We thank William Holzapfel, Shaul Hanany, Lorne Levinson, John Joseph, and Adrian Lee for many useful discussions in the design phase of this project. We are grateful to Trevor Lanting for assisting the noise measurements of the system at UC Berkeley. We acknowledge the excellent pc-board layout work of George Zizka for the Mezzanine converter board and Nick Starinski for the FPGA motherboard.

## REFERENCES

[1] M. Dobbs, N. Halverson *et. al.,* "APEX-SZ first-light and instrument status", *New Astronomy Reviews* vol. 50, 2006, pp. 960-968.
[2] John E. Ruhl *et al.*, "The South Pole Telescope", *Proc. SPIE Int. Soc. Opt. Eng*. Vol. 5543, 2004. [astro-ph/0411122]
[3] J. Fowler *et al.*, "The Atacama Cosmology Telescope project", *Proceedings of SPIE Millimeter and Submillimeter Detectors for Astronomy II*, Glasgow, Scotland, vol. 5498, 2004, pp. 1-10.
[4] H. Spieler, "Frequency Domain Multiplexing for Large Scale Bolometer Arrays", *Monterey Far-IR, Sub-mm and mm Detector Technology Workshop proceedings*, 2002, pp. 243-249.
[5] T.M. Lanting *et al.*, "Frequency domain multiplexing for bolometer arrays", *Nuclear Instruments and Methods in Physics Research A* vol. 520, 2004, pp. 548-550.
[6] J.A. Chervenak, K.D. Irwin, E.N. Grossman, J.M. Martinis, C.D. Reintsema, and M.E. Huber, "Superconducting Multiplexer for Arrays of Transition Edge Sensors", *Applied Physics Letters*, vol. 74, 1999, pp. 4043-4045.
[7] W. Holland *et al.*, "SCUBA-2: a 10,000 pixel submillimeter camera for the James Clerk Maxwell Telescope", *Proceedings of the SPIE*, vol. 6275, pp. 62751E (2006).
[8] P. Oxley *et al.*, "The EBEX Experiment", *Proc. SPIE Int. Soc. Opt. Eng*. Vol. 5543, 2004, pp. 320-331 [astro-ph/0501111v1]
[9] POLARBEAR Experiment [Online]. Available: http://bolo.berkeley.edu/polarbear/
[10] M.E. Huber et al., "DC SQUID Series Array Amplifiers with 120 MHz Bandwidth (Corrected)", IEEE Transactions on Applied Superconductivity, vol. 11 (2), 2001, pp. 4048-4053.
[11] Xilinx Logicore DDS v5.0 Product Specification, 2005 [Online]. Available: http://www.xilinx.com/ipcenter/catalog/logicore/docs/dds.pdf
[12] E. B. Hogenauer. "An economical class of digital filters for decimation and interpolation", *IEEE Transactions on Acoustics, Speech and Signal Processing*, ASSP-29(2) vol. 155, 1981.